\documentclass{article}

\usepackage[english]{babel}
\usepackage[superscript,biblabel]{cite}
\usepackage{authblk}
\usepackage[a4paper,top=3cm,bottom=2cm,left=2cm,right=2cm,marginparwidth=1.75cm]{geometry}
\usepackage{setspace}
\usepackage{amsmath}
\usepackage{amsfonts}
\usepackage{xurl}
\usepackage{graphicx}
\usepackage{booktabs}
\usepackage[colorlinks=true, allcolors=blue]{hyperref}

\date{}

\graphicspath{{Figures/}}

\usepackage{multirow}
\usepackage{subcaption} 
\usepackage{todonotes}
\usepackage{comment} 
\usepackage{siunitx}

\definecolor{rwth}{rgb}{0,0.32,0.62}
\definecolor{rwth-75}{rgb}{0.25,0.49,0.71}
\definecolor{rwth-50}{rgb}{0.55,0.73,0.89}
\definecolor{gruen}{rgb}{0.34,0.67,0.15}
\definecolor{rot}{rgb}{0.8,0.02,0.11}
\definecolor{magenta}{RGB}{227,0,102}
\definecolor{petrol}{RGB}{0,97,101}
\definecolor{violett}{RGB}{97,33,88}
\definecolor{maigrun}{RGB}{189,205,0}

\title{Graph neural networks for the prediction of molecular structure-property relationships}
\author[a]{Jan G. Rittig}
\author[b]{Qinghe Gao}
\author[c]{Manuel Dahmen}
\author[a,c]{Alexander Mitsos}
\author[b,*]{Artur M. Schweidtmann}
\affil[a]{RWTH Aachen University, Process Systems Engineering (AVT.SVT), Forckenbeckstr. 51, 52074 Aachen, Germany}
\affil[b]{Delft University of Technology, Department of Chemical Engineering, Van der Maasweg 9, Delft 2629 HZ, The Netherlands}
\affil[c]{Forschungszentrum J\"ulich GmbH, Institute of Energy and Climate Research, Energy Systems Engineering (IEK-10), Wilhelm-Johnen-Str., 52428 J\"ulich, Germany}

\affil[*]{Corresponding contributor. Email: a.schweidtmann@tudelft.com}

\begin{document}
\maketitle

\begin{abstract}

\noindent 
Molecular property prediction is of crucial importance in many disciplines such as drug discovery, molecular biology, or material and process design.
The frequently employed quantitative structure-property/activity relationships (QSPRs/QSARs) characterize molecules by descriptors which are then mapped to the properties of interest via a linear or nonlinear model.
In contrast, graph neural networks, a novel machine learning method, directly work on the molecular graph, i.e., a graph representation where atoms correspond to nodes and bonds correspond to edges.
GNNs allow to learn properties in an end-to-end fashion, thereby avoiding the need for informative descriptors as in QSPRs/QSARs.
GNNs have been shown to achieve state-of-the-art prediction performance on various property predictions tasks and represent an active field of research.
We describe the fundamentals of GNNs and demonstrate the application of GNNs via two examples for molecular property prediction.
\end{abstract}

\section{Introduction}
The estimation of molecular properties is a critical task in many disciplines such as drug discovery, molecular biology, or material and process design. 
Identification and evaluation of new molecules were originally manual, intuition-driven processes performed by scientists.
Since the 1930s, a mathematical modeling approach, the so-called quantitative structure-property relationship~(QSPR)~\cite{hammett.1935}, has provided an effective and efficient framework to screen and explore the chemical search space by computational means.
QSPRs characterize molecules with various structural, chemical, physical, and biological features, referred to as molecular descriptors, which are then mapped to a property of interest by a linear or nonlinear model.
However, the selection of informative descriptors is a difficult task that typically requires domain knowledge and expert intuition, with a bad choice leading to poor prediction performance.

Recently, advances in computational hardware and software have led to the emergence of deep learning, including novel machine learning (ML) approaches that have achieved immense success in several fields such as pattern recognition and natural language processing~\cite{LeCun2015}. 
Rather than relying on expert intuition for feature engineering, deep learning is capable of automatically learning underlying representations from data.
This has also led to a resurgence and further developments in the field of ML-based property prediction and molecule generation~\cite{Elton2019, Gilmer.2017, Wu.2018}.
In particular, graph neural networks (GNNs) have shown promising results~\cite{Gilmer.2017, Wu.2018}, as they can learn relevant features directly from the molecular graph, i.e., an intuitive representation of molecules where atoms and bonds corresponds to nodes and edges of a graph, respectively, through supervised end-to-end training.

GNNs reached state-of-the-art accuracy in predicting molecular properties, for both regression tasks, like dipole moment or enthalpy of atomization, e.g., in~\cite{Gilmer.2017, Zhang2020_MXM}, and classification tasks, such as toxicity of molecules, e.g., in~\cite{Pope2018}. 
Further refinement of GNN architectures (cf.~\cite{Wu.2021, Zhang.2022}) have led to increasing prediction accuracies on several benchmark data sets, e.g., ZINC~15~\cite{Sterling2015} or QM9~\cite{Ruddigkeit2012, Ramakrishnan2014}. 

High interest by the computer science and applied science community currently catalyzes the development of new GNN approaches and applications.
For example, we have developed higher-order GNNs~\cite{Morris.2019} and applied these to the prediction of fuel auto-ignition quality~\cite{Schweidtmann2020_GNNs}. 
GNNs have also been applied to the prediction of activity coefficients, an important problem in chemical engineering~\cite{SanchezMedina.2022}.
This rising attention shows analogies to fields in which ML models have shown remarkable achievements such as convolutional neural networks~(CNNs) for image recognition or transformer models for natural language processing.

In this manuscript, we provide an introduction to GNNs for end-to-end learning of structure-property relationships of molecules, encouraging the application of GNNs for broader use in chemical engineering.
As a starting point, we briefly outline the classical QSPR concept (Section~\ref{subsec:QSPR}) and review developments leading to ML-based molecular graph approaches (Section~\ref{subsec:NeuralMolecularGraphFingerprints}).
We then describe the concept of GNNs for molecular property prediction in detail (Section~\ref{sec:GNN4MolProp}). 
Two examples are presented applying GNNs for molecular property prediction in regression (Section~\ref{subsec:CaseStudyRegression}) and classification tasks (Section~\ref{subsec:CaseStudyClassification}).
We end the manucsript with a short conclusion, a brief overview of recent GNN developments, and an outlook (Section~\ref{sec:ConclusionOutlook}).

\section{Background}\label{sec:Background}
We first introduce QSPR modeling, an established and frequently used technique for molecular property prediction, and then present neural molecular fingerprints as a first step towards end-to-end learning of properties from molecular structure. 

\subsection{Quantitative structure-property relationships}\label{subsec:QSPR}
QSPR modeling is a two-step process consisting of (i) molecule description and (ii) property regression. 
In step (i), the structure of a molecule $m$ is utilized to compute so-called molecular descriptors $d_i$ in a mapping $D: m \mapsto \mathbf{d}$ with $\mathbf{d} = [d_1, d_2, ..., d_n]^T$. 
Note that sometimes also experimentally determined data is used as descriptor data. 
In the subsequent step (ii), the descriptors are correlated with the property $\hat{p}$ to be predicted~\cite{Katritzky.2010} with the help of a function $F(\cdot)$, i.e., 
\begin{equation}
    \hat{p} = F(\mathbf{d}) = F(D(m)) .
\end{equation}
Besides linear and nonlinear (multivariate) regression approaches, methods from ML such as random forests, support vector machines, or artificial neural networks are commonly employed for predicting the molecular property $\hat{p}$. 

Different types of molecular descriptors $\mathbf{d}$ can be distinguished according to their dimensionality\cite{Cherkasov.2014, Todeschini.2009}: 
\textit{0D-descriptors} are molecular features for which no information about the structure and connectivity between atoms is required, e.g., the count of each atom type in a molecule.
So-called \textit{1D-descriptors} describe features derived from subgroups or fragments within a molecule, e.g., the count of functional groups.
\textit{2D-descriptors} refer to the topology of a molecule and describe atoms and their bonds through a molecular graph including additional graph features like symmetry, branching, and cyclicity.
Lastly, \textit{3D-} or \textit{higher-dimensional descriptors} exploit the molecular geometry, molecular interaction fields, and time-dependent molecular dynamics. 
For a comprehensive list of molecular descriptors, the interested reader is referred to the literature~\cite{Todeschini.2000, Todeschini.2009}.

Two frequently used descriptor types are structural groups (2D) and molecular fingerprints (2D and/or 3D). 
Group contribution methods (GCMs)~\cite{Benson.1969, JOBACK.1987}, also referred to as group additivity methods, decompose the molecular structure into predefined structural groups, e.g., $>$CH-~(non-ring), =CH-~(ring), or -OH~(non-ring), with the number of occurrence of these groups constituting the molecular descriptors. 
Molecular fingerprints, on the other hand, represent molecules as vectors in which feature information about the molecular structure is stored, typically in the form of binary or integer values~\cite{Todeschini.2000, Todeschini.2009}. 
Features are, e.g., the count of substructures, similar to the GCM, or geometric distances between two atoms or structural groups~\cite{Todeschini.2009}. 

A particular challenge in QSPR modeling is the choice of suitable descriptors for a specific property prediction task at hand, as only informative descriptors may facilitate successful property prediction~\cite{Cherkasov.2014}.
Often, however, informative descriptors are not known \emph{a priori}. 
Both trial-and-error descriptor selection by the modeler and automated descriptor selection strategies can lead to chance correlations and therefore spurious results, particularly on small data sets. 
Creating a predictive QSPR model therefore may require expert intuition and must be based on a careful selection of data and algorithms for modeling and validation~\cite{Katritzky.2010, Cherkasov.2014}.

\subsection{Neural molecular fingerprints}\label{subsec:NeuralMolecularGraphFingerprints}
\begin{figure}[htbp]
    \centering
    \includegraphics[trim={8cm 6cm 8cm 6cm},clip,width=0.6\textwidth]{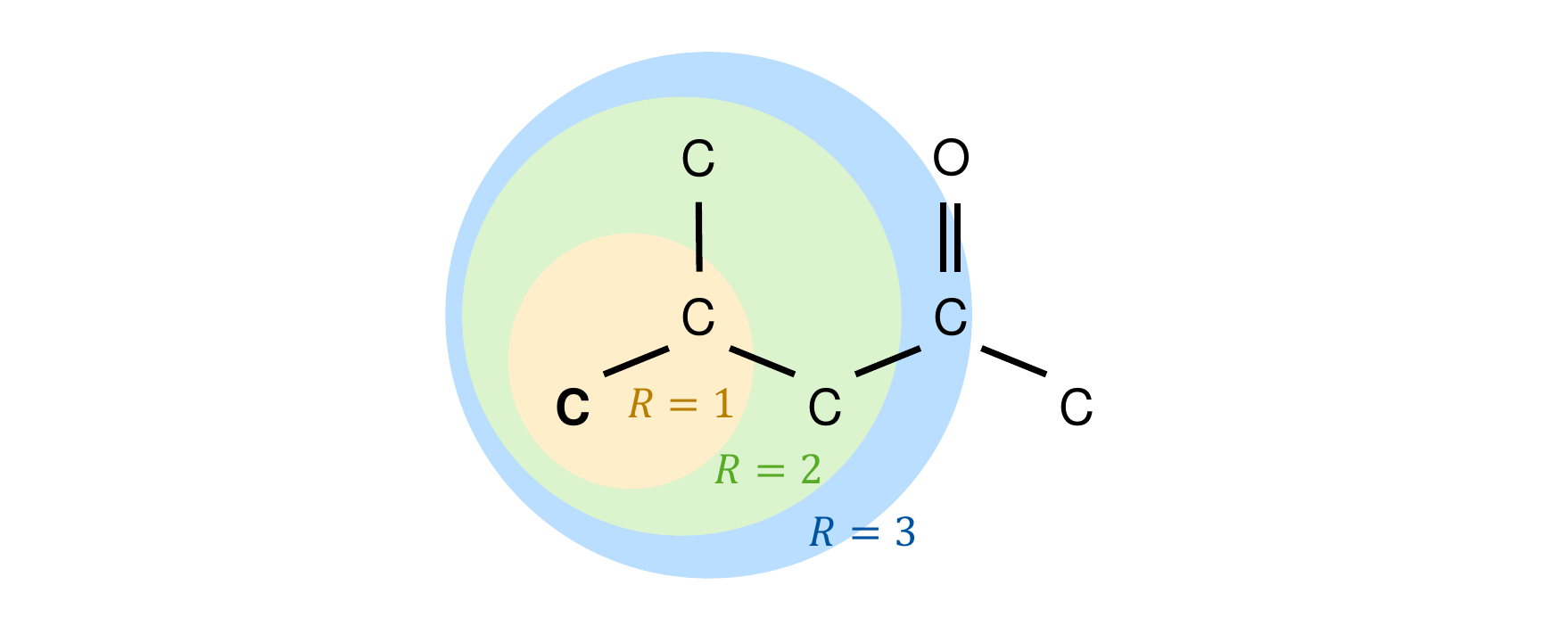}
    \caption{Neighborhood aggregation scheme for one atom (bold C atom at lower left). $R$ indicates the radius, i.e., the size of the neighborhood. Illustrative example for 4-methyl-2-pentanone.}
    \label{fig:NeighborhoodAggregation}
\end{figure}

Molecular fingerprints have long been utilized to describe molecular structures by a vector representation, with the neural molecular fingerprint being a particular type of fingerprint that bears similarities with GNNs.
Neural molecular fingerprints are based on the concept of circular molecular fingerprints~\cite{Morgan.1965, Glem.2006} that apply a neighborhood aggregation scheme to learn the local environment of an atom, often depicted as circle and illustrated in Figure~\ref{fig:NeighborhoodAggregation}. 
Each atom within a molecule is considered individually and information of its neighboring atoms is aggregated and stored in an atom feature vector~\cite{Rogers.2010}. 
Subsequently, the concatenation of the feature vector of an atom with the feature vectors of the neighboring atoms is mapped to a single-integer identifier by applying a hash function~\cite{Rogers.2010}.
The neighborhood aggregation process is repeated for a pre-defined number of iterations, referred to as radius~$R$.
As shown in Figure~\ref{fig:NeighborhoodAggregation}, the radius corresponds to the size of the neighborhood from which a node under consideration receives information.
In the first iteration, $R=1$, only information from the direct neighboring atoms can be retrieved.
In the next iteration, $R=2$, still only direct neighboring atoms pass information to each other. 
However, since all atoms have already received information about their direct neighboring atoms in the previous iteration, they can now pass on this information. 
For $R=3$, the information radius increases again, and so on.
Note that the radius is also referred to as the $R$-hop environment~(cf. Section~\ref{subsec:MessagePassingPhase}).
The learned local atom environments are then hashed to an integer value which is associated with a binary encoding in the final circular fingerprint representations.
Thus, circular fingerprint methods result in a binary fingerprint vector that encodes molecular structure information.
Popular circular fingerprints are the Morgan circular fingerprint~\cite{Morgan.1965, Glem.2006} and the ECFP~\cite{Rogers.2010}.

Duvenaud et al. (2015)~\cite{Duvenaud.2015} significantly advanced the generation of circular molecular fingerprints by modifying the neighborhood aggregation scheme to be differentiable, thereby introducing the concept of neural molecular fingerprints.
Instead of a hash function, neural molecular fingerprints apply a smooth activation function $\sigma$ (which is similar to the activation function in neural networks) for updating the atom feature vectors. 
Specifically, a softmax function is used for $\sigma$ such that the atom feature vector entries become real-valued resulting in a continuous atom vector representation.
In addition, the aggregation of neighbors is adjusted to be size-agnostic regarding the number of atoms by applying the sum operator instead of concatenating the neighbor feature vectors. 
The continuous atom feature vectors are then combined into a real-valued fingerprint vector, referred to as the neural molecular fingerprint~\cite{Duvenaud.2015}.
Further, Duvenaud et al. (2015)~\cite{Duvenaud.2015} introduced learnable parameters in the neighborhood aggregation scheme, thus allowing to train the differentiable neural molecular fingerprint with respect to specific property prediction tasks. 
Neural molecular fingerprints have been shown to outperform circular fingerprints on several prediction tasks, like solubility or drug efficacy~\cite{Duvenaud.2015}, and have paved the way for direct learning from molecular structures.

\section{Graph neural networks for molecular property prediction}\label{sec:GNN4MolProp}
GNNs are a deep learning method capable of learning properties from graph representations, thereby enabling end-to-end learning from molecular graphs to properties and eliminating the need for informative descriptors that is inherent to QSPR/QSAR modeling.
\begin{figure}[htbp]
    \centering
    \includegraphics[trim={0cm 2.5cm 0cm 2cm},clip,width=\textwidth]{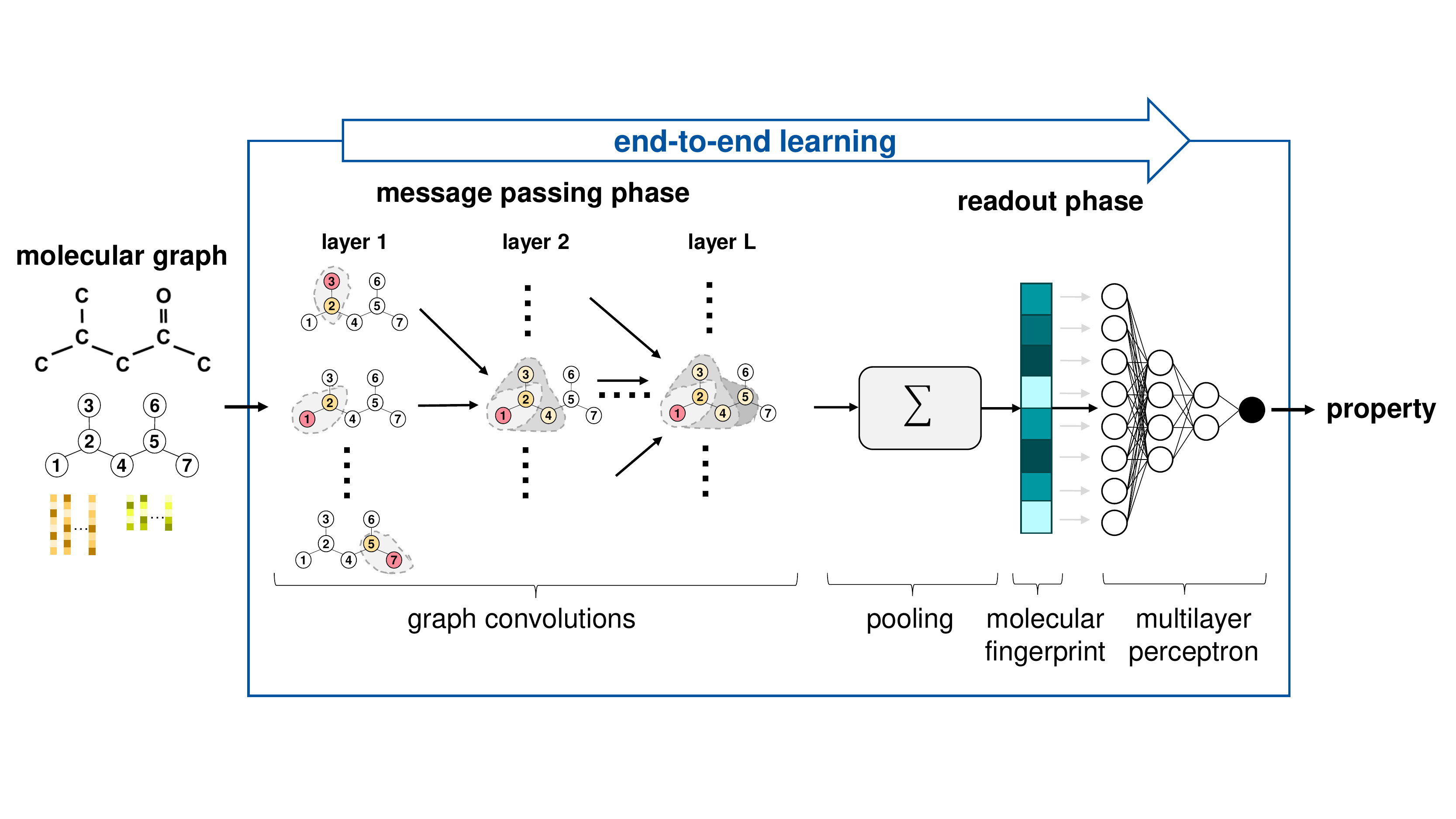}
    \caption[Overview of a graph neural network model for property prediction.]{Overview of a GNN model for property prediction, similar to the GNN we used for predicting fuel ignition qualities of hydrocarbons, cf.~\cite{Schweidtmann2020_GNNs}.}
    \label{fig:GNN_structure}
\end{figure}

As GNNs operate on graphs, a representation of nodes and edges used to model objects and their relationships, they correspond to the field of ML within the non-Euclidian data domain (e.g., grids, graphs, and groups), referred to as geometric deep learning~\cite{Bronstein.2017}. 
The concept of GNNs was introduced by Gori et al.~\cite{Gori.2005} and Scarselli et al.~\cite{Scarselli.2009}: Based on a graph as input, parametric mathematical operations are applied to learn graph characteristics that can be correlated to properties attributed to one specific node, a subregion of the graph (subgraph), or the entire graph.
As GNNs can be trained with backpropagation methods from ML, they enable end-to-end learning from the graph to the property of interest.

GNNs learn graph properties by aggregating the characteristics of local graph regions into an overall graph representation.
Their learning concept is similar to that of convolutional neural networks (CNNs), one of the most successful achievements in ML allowing to recognize objects in images~\cite{LeCun2015}.
CNNs learn by first decomposing the pixel grid into many small, local grids to extract fine-grained features and then step-wise aggregating those local grids to reassemble fine-grained features into overall characteristics of an image.
Graphs, however, are size-agnostic, i.e., the number of nodes may vary between different samples, and permutation (or isomorphism) invariant, i.e., the nodes of a graph do not have a specific order. 
GNNs are designed to address these special characteristics, i.e., they work on graphs with varying input sizes and their output does not depend on the input order of the graph nodes.
GNNs learn graph features by considering each node and its neighborhood individually, hence local graph regions, and then aggregating the single node representations into an overall graph representation.

Recently, GNNs have emerged for molecular property prediction~\cite{Kearnes.2016} from molecular graphs, a graph representation of molecules where atoms and bonds correspond to nodes and edges, respectively.
For molecular property prediction, typically, GNNs in the form of message passing neural networks (MPNNs)~\cite{Gilmer.2017} are used and are structured into two phases, the message passing phase and the readout phase, as illustrated in Figure~\ref{fig:GNN_structure}. 
In the message passing phase, the molecular structure is learned by employing so-called graph convolutions.
Graph convolutions enable each node within the graph to receive information about its neighboring nodes and edges.
In the subsequent readout phase, the learned structure information is aggregated into a molecular fingerprint vector which is then mapped to the molecular property of interest by means of, e.g., a multilayer perceptron.
All operations within the GNN structure allow for applying backpropagation, enabling end-to-end learning of molecular properties from the molecular graph.

We outline the concept of molecular graphs (Section~\ref{subsec:MolecularGraph}), the message passing phase (Section~\ref{subsec:MessagePassingPhase}), and the readout phase (Section~\ref{subsec:ReadoutPhase}) in more detail, and then describe the end-to-end learning of molecular properties (Section~\ref{subsec:EndToEndLearning}).

\subsection{Molecular graph}\label{subsec:MolecularGraph}
We denote the graph representation of a molecule $m$ with nodes as atoms, also referred to as vertices $v \in V$, and edges as bonds $e_{vw} \in E$ connecting two nodes $v$ and $w$ by $G(m) = \{V, E\}$.
\begin{figure}
    \centering
    \includegraphics[trim={0cm 2.5cm 0cm 3cm},clip,width=\textwidth]{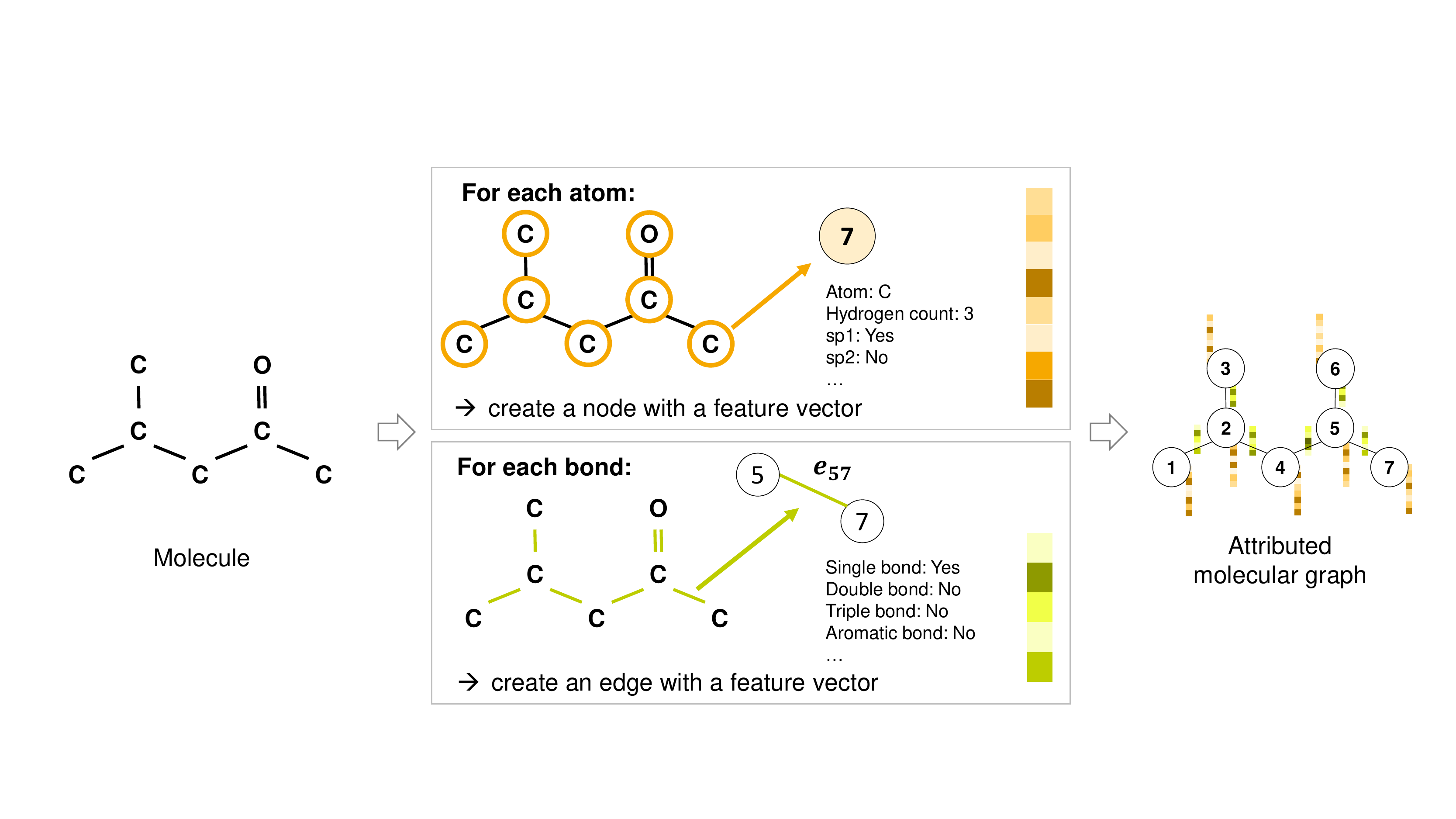}
    \caption{Generation of an attributed molecular graph, illustrative example for 4-methyl-2-pentanone.}
    \label{fig:MolecularGraph}
\end{figure}

Each node and each edge of the graph $G$ is typically associated with a feature vector that stores atom-/bond-specific information, e.g., the atom type or the bond type.
We denote the feature vector of a node $v$ with $\mathbf{f}^v \in F^V$ and the feature vector of an edge $e_{vw}$ with $\mathbf{f}^{e_{vw}} \in F^E$.
Typical features are given in Table~\ref{tab:AtomFeatures} for atoms and in Table~\ref{tab:BondFeatures} for bonds.
The feature-enriched graph of a molecule $m$ is then denoted by $G(m) = \{V, E, F^V, F^E\}$, often referred to as attributed molecular graph (cf. Figure~\ref{fig:MolecularGraph}).

\begin{table*}[htbp]
	\centering
	\caption{Atom features for initial node feature vector~\cite{Gilmer.2017, Coley.2017, Yang.2019, Schweidtmann2020_GNNs}. Features are typically encoded as one-hot vectors.}
	\begin{tabular}{ll}
		\toprule
		\textbf{Feature} & \textbf{Description/exemplary values} \\
		\midrule
		atom type & type of atom (C, O, S, F) \\
		is in ring & whether the atom is part of a ring \\
		is aromatic & whether the atom is part of an aromatic system \\
		hybridization & sp, sp2, sp3, sp3d, or sp3d2 \\
		charge & formal charge of the atom \\
		\# bonds & number of bonds the atom is involved in \\
		\# Hs & number of bonded hydrogen atoms \\
		\bottomrule
	\end{tabular}
	\label{tab:AtomFeatures}
\end{table*}

\begin{table*}[htbp]
	\centering
	\caption{Bond features for initial edge feature vector~\cite{Gilmer.2017, Coley.2017, Yang.2019, Schweidtmann2020_GNNs}. Features are typically encoded as one-hot vectors.}
	\begin{tabular}{ll}
		\toprule
		\textbf{Feature} & \textbf{Description/exemplary values} \\
		\midrule
		bond type & single, double, triple, or aromatic \\
		conjugated & whether the bond is conjugated \\
		is in ring & whether the bond is part of a ring \\
		\bottomrule
	\end{tabular}
	\label{tab:BondFeatures}
\end{table*}

Note that the representation of molecules as graphs is a simplification as it omits, e.g., 3D information.
Whereas node/edge features can include additional information, e.g., distances or angles between atoms, some characteristics of molecules such as information about stereochemistry or molecule conformation are arguably difficult to capture with a 2D representation such as graphs.

The generation of molecular graphs can be automated by using open-source tools, e.g., RDKit~\cite{rdkit}, that can also calculate typical node and edge features.
Methods from graph-based ML such as GNNs can then be applied to the generated attributed molecular graphs.

\subsection{Message passing}\label{subsec:MessagePassingPhase}
\begin{figure}
    \centering
    \includegraphics[trim={0cm 0cm 0cm 0cm},clip,width=\textwidth]{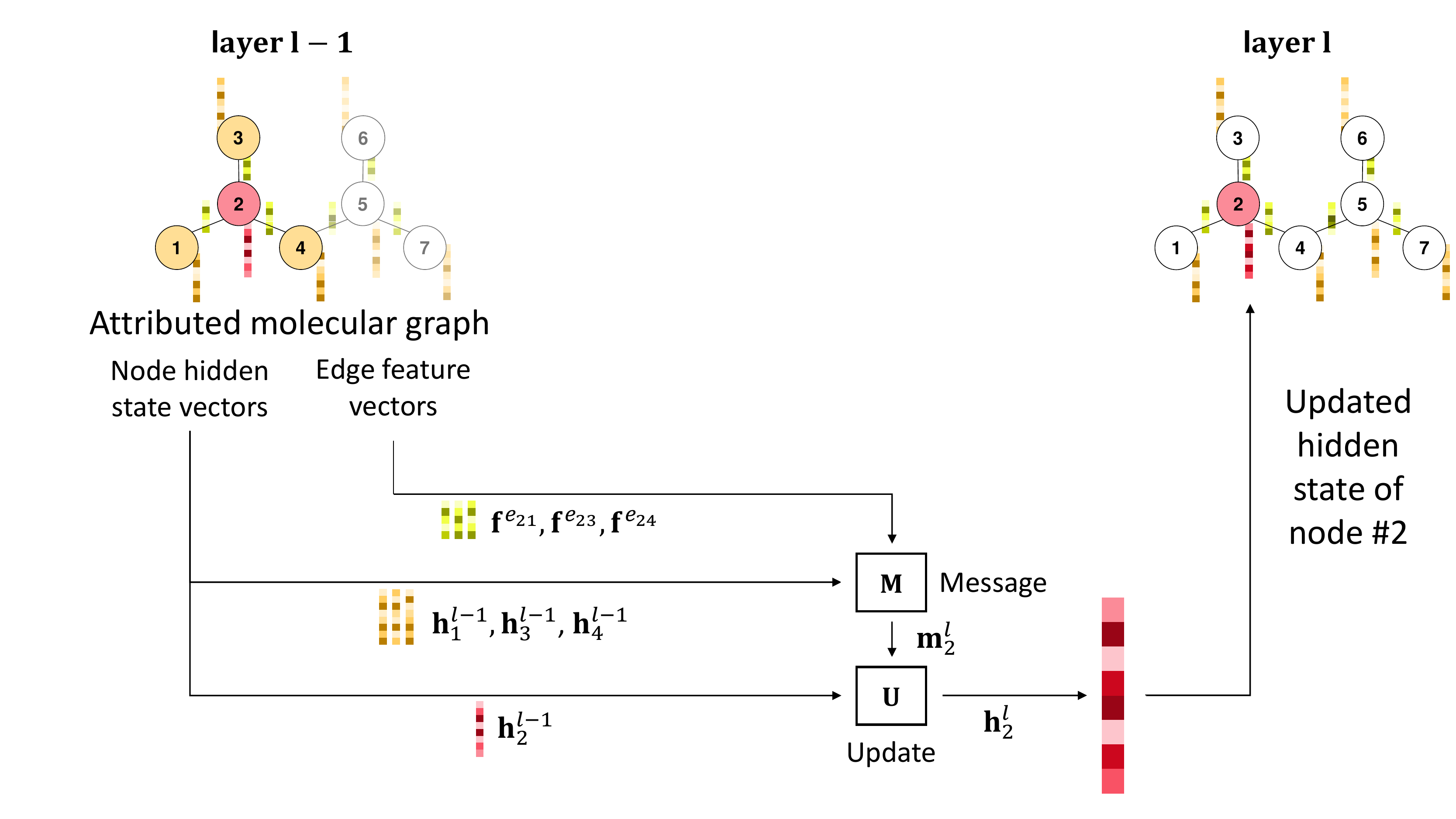}
    \caption{Information exchange between nodes within an edge-conditioned graph convolutional layer in the message passing phase of a graph neural network, illustrated update step for node \#2, adapted from our publication~\cite{Schweidtmann2020_GNNs}.}
    \label{fig:MessagePassing}
\end{figure}

In the message passing phase, structural information about a molecule is extracted from the molecular graph through graph convolutions. 
Graph convolutions enable nodes to exchange information with their neighboring nodes, i.e., messages are passed along edges (cf. Figure~\ref{fig:MessagePassing}).
Specifically, graph convolutions combine the feature vector of each node with the features vectors of the corresponding neighboring nodes and edges, i.e., messages containing neighborhood information are passed between nodes (cf. neighborhood aggregation in Section~\ref{subsec:NeuralMolecularGraphFingerprints}). 
Stacking a total of $L$ graph convolutional layers together, each node receives information about an environment with a radius of $L$ nodes.
The general mathematical form of a graph convolutional (or message passing) layer is described in the following.

In every graph convolutional layer $l$, each node is represented by a latent or hidden state $\mathbf{h}_v^l$. 
The hidden state of a node is initialized by its feature vector from the attributed molecular graph, $\mathbf{h}_v^0 = f^v$ with $l=0$ (cf. Section~\ref{subsec:MolecularGraph}). 
The molecular graph then traverses the individual graph convolutional layers $l \in \{1, 2, ..., L\}$. 
Within a graph convolutional layer $l$, the hidden states of the nodes within the molecular graph are updated based on the hidden states of the respective neighborhood nodes and the features of the associated edges.
The update step can be denoted by the general form of~\cite{Hamilton.2017_Inductive, Gilmer.2017}
\begin{equation}
    \mathbf{m}_v^l = A_{l} (\{ M_{l}(\mathbf{h}_v^{l-1}, \mathbf{h}_w^{l-1}, \mathbf{f}^{e_{vw}})) \mid w \in N(v) \}),
    \label{eq-update_message}
\end{equation}
\begin{equation}
    \mathbf{h}_v^l = U_l(\mathbf{h}_v^{l-1}, \mathbf{m}_v^{l}) \label{eq-update_rule} .
\end{equation}
The function $M_l$ denotes the message function mapping the information, the so-called \emph{message}, from one of its neighbor nodes $w$ to node $v$, with $w \in N(v)$. 
The message is a function of the preceding hidden states $\mathbf{h}_v^{l-1}$, $\mathbf{h}_w^{l-1}$, and edge $e_{vw}$ with the associated edge feature vector $\mathbf{f}^{e_{vw}}$.%
\footnote{
Note that typically the node states are updated and the edge features remain unchanged.
However, there also exists GNN architectures where the edge information changes during the update process or even the whole graph convolutional layer update process is based on messages associated with edges rather than nodes, see, e.g., ~\cite{Yang.2019}.
}
The messages from all neighbors of node $v$ are then aggregated in $\mathbf{m}_v^l$ by applying the aggregation function $A_{l}$. 
These neighbor messages together with the preceding hidden state of the node $v$ itself, $\mathbf{h}_v^{l-1}$, are finally combined in the update function $U_l$ to update the hidden state of node $v$. 
Consequently, the hidden state of a node in layer $l$ depends on its own previous hidden state, the previous hidden states of its neighboring nodes, and the associated edges.
Note that the size of the hidden state vector is a hyperparameter and may vary for different layers. 

The $L$-hop local environment of a node is learned in the message passing phase which builds on neighborhood aggregation (cf. Section~\ref{subsec:NeuralMolecularGraphFingerprints}).
By stacking multiple graph convolutional layers $L$ together, the information of all neighboring nodes within a distance of $L$ nodes is passed to a node, hence the so-called $L$-hop local environment of each node is learned.
Accordingly, the number of layers $L$ is also referred to as the \emph{depth} or \emph{radius} of the network, i.e., the local environment of a node reachable with $L$ so-called hops is extracted. 
Various approaches exist for the aggregation, the message function  $M_{l}$, and the update function $U_l$ in graph convolutional layers.

For aggregating the messages passed to a node within a graph convolutional layer by the function $A_{l}$, different operators can be used, e.g., mean or max, as well as LSTM aggregators~\cite{Hamilton.2017_Inductive}.
The sum aggregator has been shown to be more powerful than the mean or max aggregator in terms of distinguishing different neighborhood structures of a node~\cite{Xu.2018}.
Thus, typically the sum aggregator is used for molecular property prediction, i.e., 
\begin{equation}
    \mathbf{m}_v^l = \sum_{w \in N(v)} M_{l}(\mathbf{h}_v^{l-1}, \mathbf{h}_w^{l-1}, \mathbf{f}^{e_{vw}})
    \label{eq-update_message-sum}.
\end{equation}
For the message function and the update function in a graph convolutional layer, a wide variety of formulations has been proposed, e.g., graph convolutional network (GCN)~\cite{Kipf2016}, graph attention network (GAT)~\cite{Velickovic.2017}, GraphSAGE~\cite{Hamilton.2017_Inductive}, and higher-order graph convolutions~\cite{Morris.2019}.
Further, graph convolutional layers can be combined with a gated recurrent unit (GRU)~\cite{Cho.03.06.2014, Li.17.11.2015, Gilmer.2017}.
Here, we present two graph convolutional approaches: Graph convolutions based only on node features, which we refer to as basic graph convolution, and edge-conditioned graph convolutions, which also take into account edge features.
The interested reader is referred to the reviews in~\cite{Gilmer.2017, Wu.2021, Zhang.2022} for further graph convolutional approaches.

\paragraph{Basic graph convolution.}\label{subsubsec:StandGraphConv}
A basic form of graph convolution is a linear transformation of the node states by using a parameter matrix $W_1^l$ to transform the state of a considered node $v$ and the parameter matrix $W_2^l$ to incorporate messages passed from the neighbors of node $v$. 
Specifically, the parameter matrix $W_1^l$ is used to transform the hidden state vector of node $v$ from the previous layer, $\mathbf{h}_v^{l-1}$, whereas the parameter matrix $W_2^l$ is multiplied by the hidden state vectors of the neighboring nodes, $\mathbf{h}_w^{l-1}$.
Note that edge features are omitted here.
The results of the respective multiplications are summed up and transformed by an activation function $\sigma$, e.g., ReLU, resulting in the updated hidden state of the node $\mathbf{h}_v^l$, i.e.,
\begin{equation}
    \mathbf{h}_v^l = \sigma(W_1^l \cdot \mathbf{h}_v^{l-1} + \sum_{w \in N(v)} W_2^l \cdot \mathbf{h}_w^{l-1}). \label{eq-sum_update_rule}
\end{equation}

\paragraph{Edge-conditioned graph convolution.}\label{subsubsec:EdgeGraphConv}
Molecular graphs are typically augmented with edge features, e.g., whether a bond between two atoms is a single, double, or triple bond (cf. Section~\ref{subsec:MolecularGraph}). 
To include edge features in the graph convolutions, edge-conditioned graph convolutions have been proposed~\cite{Simonovsky.10.04.2017}.
The update step in an edge-conditioned convolution is illustrated in Figure~\ref{fig:MessagePassing}.
For edge-conditioned graph convolutions the message function in Equation \eqref{eq-update_message} is defined by
\begin{equation}
     M_l = \text{ANN}_{W_E^l}(\mathbf{f}^{e_{vw}}) \cdot \mathbf{h}_w^{l-1} \quad \text{with} \quad w \in N(v) \label{eq-edge-filtering_message-function},
\end{equation}
where $ANN$ denotes an artificial neural network mapping the edge features to a parameter matrix that is multiplied with the hidden states of the neighbors of node $v$, $\text{ANN}_{W_E^l}:~\mathbf{f}^{e_{vw}}~\mapsto~W_E^l$.
The hidden state of the node $v$ is again multiplied by a parameter matrix $W_V^l$, similarly to the standard graph convolution.
The edge-conditioned message function combined with the sum aggregator then results in the following update process:
\begin{equation}
    \mathbf{h}_v^l = \sigma(W_V^l \cdot \mathbf{h}_v^{l-1} + \sum_{w \in N(v)} \text{ANN}_{W_E^l}(\mathbf{f}^{e_{vw}}) \cdot \mathbf{h}_w^{l-1}) \label{eq-edge-filtering_update}
\end{equation}

\subsection{Readout}\label{subsec:ReadoutPhase}
In the \emph{readout phase}, the structure information learned during message passing and stored in the hidden nodes is first aggregated into a molecular fingerprint vector. 
We denote the aggregation by the pooling function $P$, i.e., 
\begin{equation}
    \mathbf{h}_{FP} = P(\{\mathbf{h}_v^L \mid v \in V\}) \label{eq-edge-pool}, 
\end{equation}

\begin{figure}
    \centering
    \includegraphics[trim={6cm 7cm 8cm 6cm},clip,width=\textwidth]{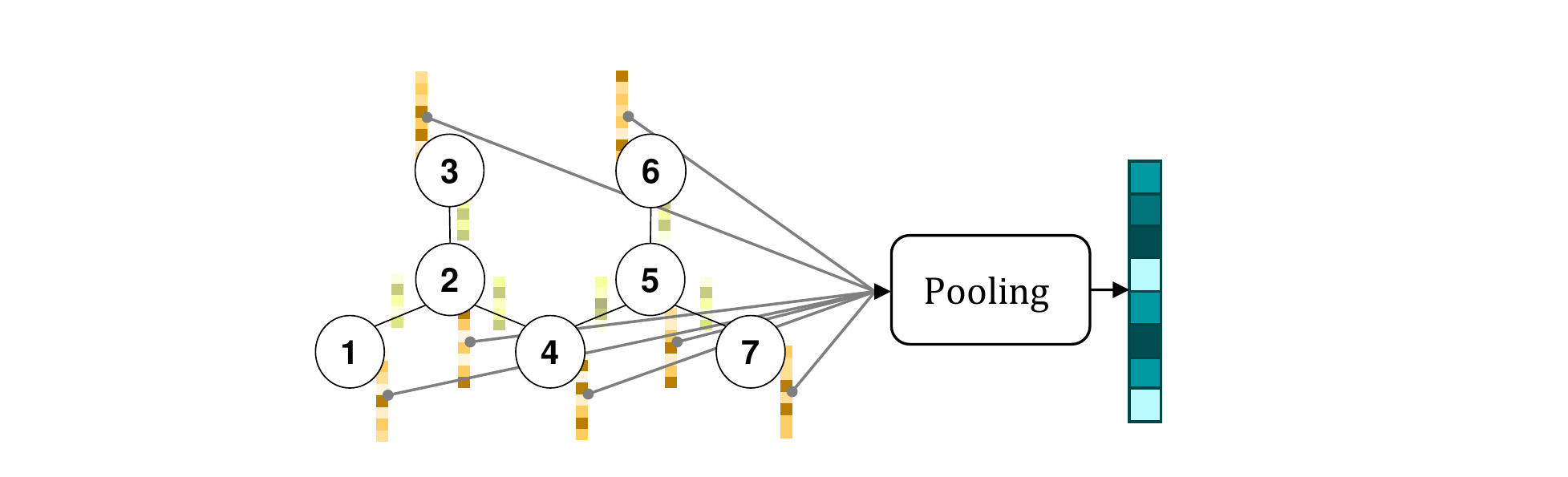}
    \caption{Pooling step in a graph neural network, illustrative example for 4-methyl-2-pentanone.}
    \label{fig:Pooling}
\end{figure}

\noindent that combines the hidden states of the nodes of the last convolutional layer $L$ into the molecular fingerprint $\mathbf{h}_{FP}$ (cf. Figure~\ref{fig:Pooling}), a continuous molecular vector representation.
The size of the molecular fingerprint is implicitly defined by the size of the node hidden state vectors in the last graph convolutional layer which is a hyperparameter (cf. Section~\ref{subsec:MessagePassingPhase}).

Different pooling functions exist and range from simple mathematical operations, e.g., mean, sum, or max, to advanced neural network approaches~\cite{Wu.2021}.
A crucial condition for these pooling functions is that the function should be invariant to the order of nodes, hence permutation/isomorphism invariant~\cite{Xu.2018, Zhang.2022}, and size-agnostic, i.e., applicable to a varying number of nodes. 
Flexible pooling functions have been developed in recent years~\cite{Vinyals.11192015, Zhang.2018, Ying.2018}, one popular example being the set2set approach~\cite{Vinyals.11192015}. 
The set2set model is designed to operate on input sets with an invariance for the order of the objects~\cite{Vinyals.11192015, Gilmer.2017} and employs an
attention mechanism including a LSTM~\cite{Vinyals.11192015} that yields the molecular fingerprint after $T$ steps, i.e., $\mathbf{h}_{FP} = \mathbf{q}_{t=T}^{*}$, by executing the iterative scheme
\begin{equation}
	\begin{gathered}
		\mathbf{q}_{t} = \text{LSTM}\left(\mathbf{q}_{t-1}^{*}\right), \\
		\mathbf{a}_{v,t} = \text{softmax}\left( \mathbf{h}_{v}^{L} \cdot \mathbf{q}_{t} \right), \\
		\mathbf{r}_{t} = \sum_{v} \mathbf{a}_{v,t} \cdot \mathbf{h}_{v}^{L}, \\
		\mathbf{q}_{t}^{*} = \mathbf{q}_{t} \mathbin\Vert \mathbf{r}_{t}, \label{eq-set2set} \\
	\end{gathered}
\end{equation}
where $\mathbf{q}_{t}$ is a query vector for iteration $t$ providing information from the last iteration by an LSTM, $\mathbf{a}_{v,t}$ is an attention vector yielding the attention of a node $v$ by applying the softmax function, $\mathbf{r}_{t}$ is the attention readout vector calculated by the attention-weighted sum of the single node hidden states, and $\mathbf{q}_{t}$ is a concatenation ($\mathbin\Vert$) of the query vector and the attention readout vector. 
The vector $\mathbf{q}_{t-1}^{*}$ can be initialized with the zero-vector, i.e., $\mathbf{q}_{-1}^{*}=\mathbf{0}$. 

Finally, the molecular fingerprint $\mathbf{h}_{FP}$ must be mapped to the property of interest. 
Here, standard regression methods can be employed. 
In GNNs, typically, a feedforward neural network, a multilayer perceptron~(MLP), is used to compute the molecular property $\hat{p}$ as a function of the molecular fingerprint vector, i.e.,
\begin{equation}
    \hat{p} =  \text{MLP}(\mathbf{h}_{FP}).
\end{equation}

\subsection{End-to-end learning}\label{subsec:EndToEndLearning}
We denote the entire feedforward function $f_{GNN}$ of the GNN for prediction of a molecular property $\hat{p}$ from the molecular graph $G(m)$ by
\begin{equation}
    f_{GNN}: G(m) \mapsto \hat{p}.
\end{equation}

All mathematical operators used in the GNN for the mapping $f_{GNN}$ can be integrated into the backpropagation scheme of ML models. 
In particular, backpropagation can be used to learn the parameters of all three functions in the message passing phase (cf. Section~\ref{subsec:MessagePassingPhase}), i.e., the message function $M_l$, the aggregation function, and the update function $U_l$.
Likewise, the pooling function and the feedforward ANN in the readout phase (cf. Section~\ref{subsec:ReadoutPhase}) allow for backpropagation.
Thus, the whole GNN structure can be trained with backpropagation in a supervised setup~\cite{Gilmer.2017, Hamilton.2017_Inductive}. 
Therefore, GNNs are able to learn in an end-to-end manner, from the molecular graph to the property of interest.
As the molecular fingerprint automatically adapts during the GNN training to capture the structural information that is important for the prediction task at hand~\cite{Coley2017}, GNNs circumvent the definition and selection of informative molecular descriptors as it is required in QSPR/QSAR modeling. 

\section{Numerical examples}\label{sec:ResultsDiscussion}
To demonstrate typical workflows of applying GNNs for molecular property prediction, we consider two examples and focus on data pre-processing, model setup and training, and result analysis.
Both examples are implemented in Python using the open-source extensions PyG (PyTorch Geometric)~\cite{Fey2019_PyGeo}, a package for geometric ML which includes a wide range of GNN architectures, and RDKit~\cite{rdkit}, a chemoinformatics package (cf. Section~\ref{subsec:MolecularGraph}).
The first example is a regression task, i.e., a real-valued property of a molecule, herein the boiling point, shall be predicted. 
The second example is a classification problem. 
Specifically, a molecule shall be categorized as biodegradable or non-biodegradable.
The examples show that GNN models with standard architecture can provide high prediction accuracies, enabling fast and accurate predictions that can, for instance, be used in computer-aided molecular design.

\subsection{Regression example: Boiling point prediction}\label{subsec:CaseStudyRegression}

The normal boiling point is often used to assess the general suitability of a chemical component as a raw/working material, a reactant, or a product within chemical process design.
For novel compounds, experimentally determined boiling points are often not readily available.
In this example, we demonstrate the development of a GNN for normal boiling point prediction. 

\paragraph{Data set and preprocessing} \mbox{}\vspace{-0.5cm} \\

\noindent We use the normal boiling point data set provided by Alsheri et al.~\cite{Alshehri.2021} which contains 5,276 boiling points for 5,089 molecules of various classes, i.e., pure hydrocarbons, hydrocarbons with additional atoms such as oxygen, nitrogen, or fluorine, as well as multi-functional molecules.
Initially, we randomly select 10\,\% of the data set for testing, i.e., 509 molecules are set aside. 
Note that the test set is kept unchanged in the following. 
The remaining 4,749 data points are split randomly into 90\,\% training set and 10\,\% validation set.
We convert the molecules provided as SMILES strings to attributed molecular graphs with the atom and bond features provided in Table~\ref{tab:AtomFeatures} and Table~\ref{tab:BondFeatures}, respectively, with the help of RDKit~\cite{rdkit}.
Note that we extend the atom features stated in Table~\ref{tab:AtomFeatures} to the atom types occurring in the data set (C, O, N, F, S, Cl, P, I, Br, Si) and exclude atom features that do not occur or do not vary, i.e., sp3d and atom charge, resulting in 20 atom features and 6 bond features in total.
We standardize the boiling point values to a Gaussian with a zero mean and standard deviation of one (Z-score normalization) for training purposes.

\paragraph{Model setup and training} \mbox{}\vspace{-0.5cm} \\

\noindent We use a GNN architecture with three edge-conditioned graph convolutional layers in the message passing phase and sum pooling followed by a three-layer MLP in the readout phase, cf. Section~\ref{sec:GNN4MolProp}.
For the graph convolutions, the hidden state vectors are set to a dimension of 64 yielding a 64-dimensional molecular fingerprint.
The edge-MLPs within the graph convolutions have three layers with 6, 128, and 4092 (=64$^\text{2}$) neurons, respectively. 
The three layers in the MLP for the readout have 64, 32, and 1 neurons, with the last neuron yielding the boiling point prediction.

To train the GNN model, we minimize the loss, i.e., the mean squared error between boiling point predictions and labels, by using the Adam optimizer. 
We train the model for a maximum of 300 epochs and apply early stopping with a patience of 25 epochs, i.e., the training is stopped early if the validation mean absolute error does not decrease for 25 consecutive epochs.
The learning rate is initialized with a value of 0.001 and decayed with a factor of 0.8 if the validation error does not decrease for 3 consecutive epochs (patience).
We further use a batch size of 8.
In addition to training the single GNN prediction model, we employ ensemble learning which means that we train 40 GNN models with different random splits of the data set into training and validation sets and average the predictions of the 40 GNNs to compute the final boiling point prediction.

\paragraph{Prediction results} \mbox{}\vspace{-0.5cm} \\

\noindent Table~\ref{tab:Prediction_Tb} shows the boiling point T$_\text{b}$ prediction accuracies achieved by the GNN.
Note that we report the average accuracy and corresponding standard deviation with respect to training, validation, and test set data.
As the ensemble GNN contains 40 models with varying training and validation splits, we state a single aggregated training and validation set accuracy. 
For the single GNN model, the statistics are computed over 40 single models.

The mean absolute error (MAE) of the single GNN models on the test set is 14.4 \si{K} which corresponds to a mean absolute percentage error (MAPE) of about 3.2\,\% and a coefficient of determination R$^2$ of 0.86.
Comparing the prediction quality of the single GNN models to the ensemble GNN, we observe a clear advantage for the ensemble, i.e., an increase in accuracy with respect to both the combined training/validation set and the test set.
Specifically, the MAE on the test set is reduced by 2.5 \si{K} to 11.9 \si{K}, resulting in an R$^2$ of 0.90.
The increased accuracy results from over- and underpredictions of single GNNs for the same molecule being partly balanced out by taking the average prediction.

\begin{table}[htbp]
    \centering
    \resizebox{\linewidth}{!}{%
    \begin{tabular}{lllll}
        \toprule
        \multicolumn{1}{c}{\multirow{2}[3]{*}{\shortstack[c]{Model setup}}} & \multicolumn{4}{c}{Boiling point, T$_\text{b}$} \\
        \cmidrule(lr){2-5}
        \multicolumn{1}{c}{}   & \multicolumn{1}{c}{MAE/\si{K}} & \multicolumn{1}{c}{R$^2$} & \multicolumn{1}{c}{MAPE/\%} & \multicolumn{1}{c}{MAXPE/\%} \\
        \midrule
        single GNN (training set) & 8.7 $\pm$ 2.1 & 0.93 $\pm$ 0.04 & 1.9 $\pm$ 0.4 & 48.4 $\pm$ 13.2 \\
        single GNN (validation set) & 13.3 $\pm$ 1.8 & 0.85 $\pm$ 0.13 & 2.9 $\pm$ 0.3 & 33.7 $\pm$ 23.8 \\
        single GNN (test set) & 14.4 $\pm$ 1.3 & 0.86 $\pm$ 0.04 & 3.2 $\pm$ 0.3 & 41.9 $\pm$ 4.4 \\
        ensemble of 40 GNNs (training/validation set) & 7.1 & 0.94 & 1.5 & 42.5 \\
        ensemble of 40 GNNs (test set) & 11.9 & 0.90 & 2.6 & 41.9 \\
        \bottomrule
    \end{tabular}
    }
    \caption{Model accuracies for boiling point (T$_\text{b}$) prediction: mean absolute error (MAE) in Kelvin (\si{K}), coefficient of determination (R$^2$), mean absolute percentage error (MAPE), and maximum absolute percentage error (MAXPE). For the single GNN model, the statistics are computed over 40 single models.}
    \label{tab:Prediction_Tb}
\end{table}

\begin{figure}[htbp]
    \centering
    \captionsetup[subfigure]{justification=centering}
	\begin{subfigure}[c]{0.5\textwidth}
		\centering
		\includegraphics[width=\textwidth]{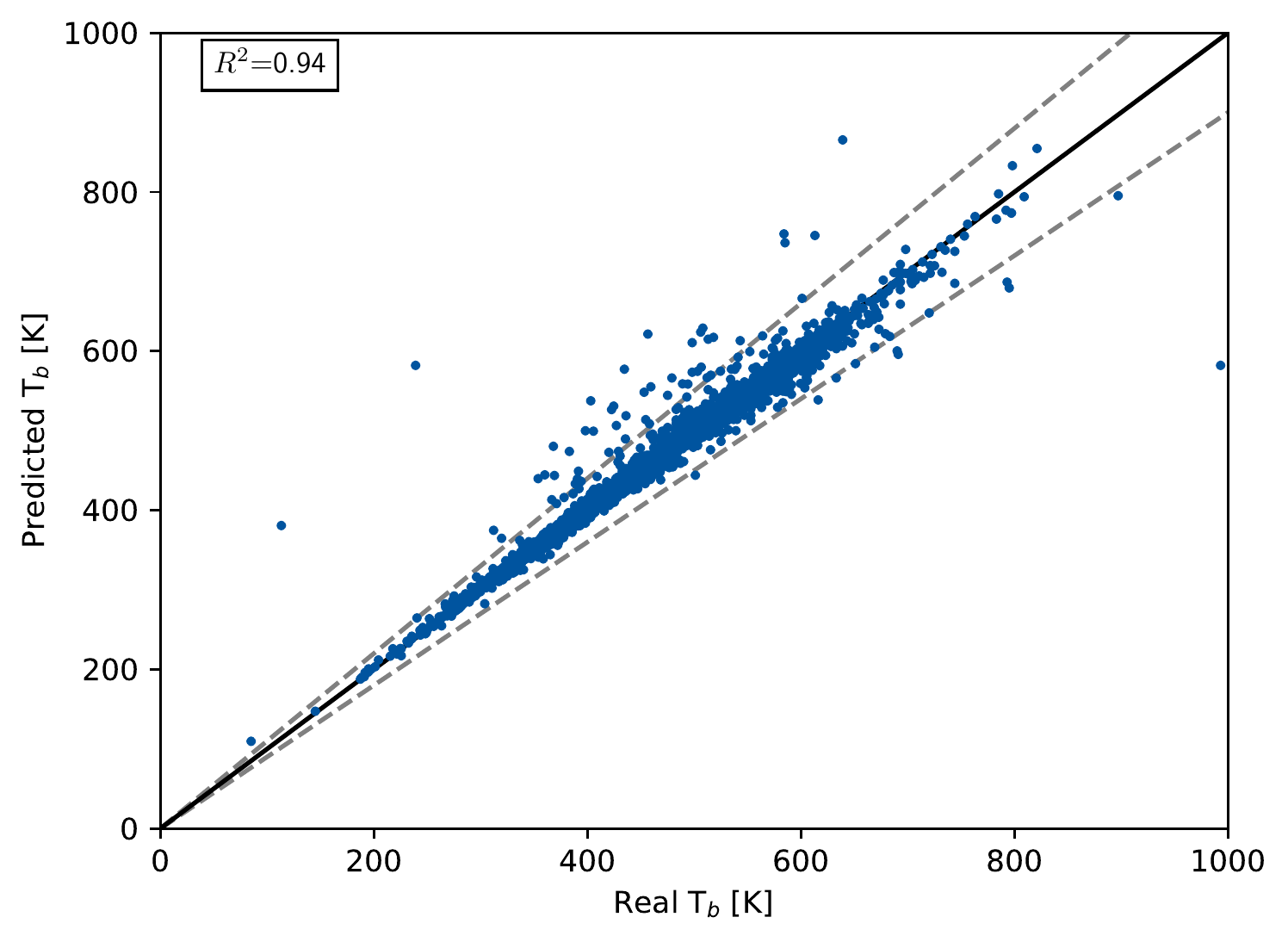}
		\subcaption{Training/validation set}
	\end{subfigure}%
	\begin{subfigure}[c]{0.5\textwidth}
		\centering
		\includegraphics[width=\textwidth]{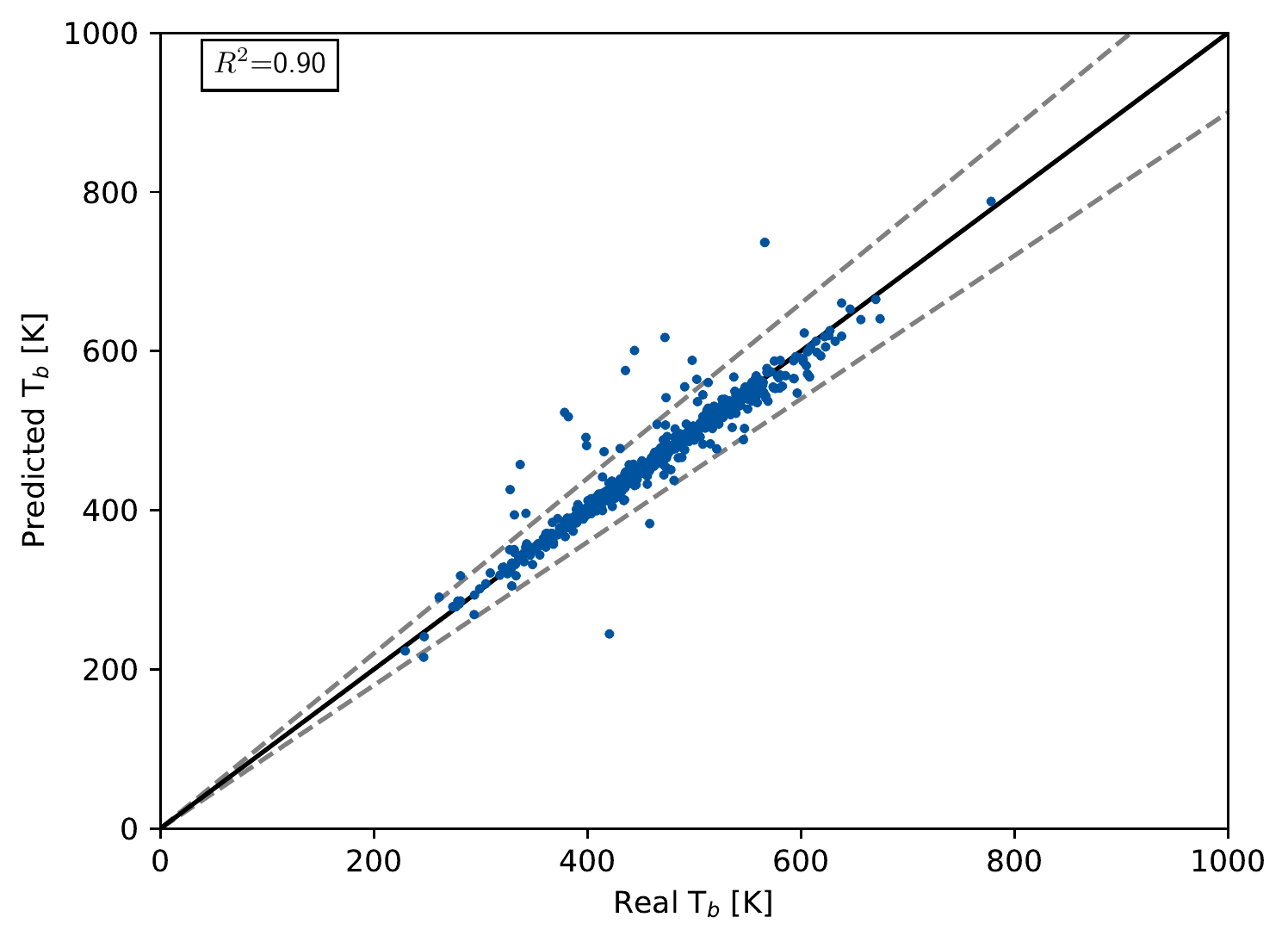}
		\subcaption{Test set}
	\end{subfigure}
	\caption{Parity plots for GNN ensemble. Grey dashed lines indicate $\pm$ 10\,\% error. Note that one sample within the training/validation set with a measured boiling point of about 2500 \si{K} is not shown for better visualization.}
	\label{fig:Tb_ParityPlots}
\end{figure}

Experimental boiling points and predictions by the ensemble for the molecules within the training/validation set and the test set are visualized in the parity plots in Figure~\ref{fig:Tb_ParityPlots}.
Most predictions deviate less than $\pm$ 10\,\% from the actual values, indicated by the grey dashed lines (Figure~\ref{fig:Tb_ParityPlots}).
More specifically, for the training/validation set about 99\,\% and for the test set about 95\,\% of the data points are predicted with an absolute error below 10\,\%.
For both sets some outliers can be observed. 
Nevertheless, the total number of outliers is low and the residuals are approximately normally distributed.
We thus find an overall high prediction quality. 

\subsection{Classification example: Biodegradability prediction}\label{subsec:CaseStudyClassification}
The assessment of properties that indicate environmental impacts and hazards of chemicals is a critical step in the design of sustainable products and processes.
A property of paramount importance here is the persistence of chemicals in the environment, as the accumulation of persistent materials may exhibit long-term toxic effects.
This classification example aims at categorizing molecules into readily biodegradable or non-readily biodegradable substances by means of GNN modeling.

\paragraph{Data set and preprocessing} \mbox{}\vspace{-0.5cm} \\

\noindent For binary biodegradability classification, Mansouri et al.~\cite{Mansouri2013} provided the ``UCI QSAR Biodegradation Data-Set'', a data set consisting of 1,725 molecules with a binary biodegradability label.
The class of readily biodegradable compounds contains 547 molecules, whereas 1,178 molecules are considered non-readily biodegradable. 
We use the pre-defined test set with 670 molecules from Mansouri et al.~\cite{Mansouri2013} and the remaining 1,055 data points with a random split into 90\,\% training and 10\,\% validation data for training purposes.
Again, attributed molecular graphs with atom and bond features as provided in Table~\ref{tab:AtomFeatures} and Table~\ref{tab:BondFeatures}, respectively, are inputted to the GNN model, where we extend the atom features to the atom types occurring in the data set (C, O, N, F, Si, S, P, Cl, Ti, Cu, Br, Sn, I, Bi) and exclude one atom feature not occurring in the data, i.e., sp3d, yielding 27 atom features and 6 bond features in total.

\paragraph{Model setup and training} \mbox{}\vspace{-0.5cm} \\

\noindent The GNN model architecture is similar to that of the regression example described above. 
We employ three edge-conditioned graph convolutional layers in the message passing phase and apply sum pooling with a three-layer MLP for the readout. 
The hidden state vectors of the graph convolutions and molecular fingerprint have 64 dimensions; the edge-MLPs have three layers with, respectively, 6, 128, and 4092 (=64$^\text{2}$) neurons. 
The three layers in the readout-MLP have 64, 32, and 1 neurons.
In case of binary classification, the readout-MLP predicts the probability of a positive class, herein, readily biodegradable, by applying a sigmoid function in the last layer.
We classify a molecule as readily biodegradable if the probability is higher than a threshold value of 0.5, a hyperparameter of the model.

To train the GNN model, we use the Adam optimizer with the binary cross-entropy loss function, a batch size of 8, an initial learning rate of 0.001, and learning rate decay with a factor of 0.8 and a patience of 3 epochs.
The training ends either after 300 epochs or when the validation cross-entropy loss does not decrease for 25 consecutive epochs (early stopping).
We again perform single GNN modeling and ensemble learning. 
In the latter case, we train 40 GNN models with different random data set splits and average the property predictions, i.e., a molecule is classified as biodegradable if the average probability prediction of the 40 models yields a value higher than 0.5.

\paragraph{Prediction results} \mbox{}\vspace{-0.5cm} \\

\noindent Table~\ref{tab:Prediction_BioDeg} shows several performance indicators of the GNN models.
The average performance and corresponding standard deviation for 40 single GNN models with respect to training, validation, and test sets are reported.
For the GNN ensemble, the performance on the aggregated training/validation set (cf. Section~\ref{subsec:CaseStudyRegression}) and the test set is indicated.

\begin{table}[htb]
    \centering
    \resizebox{\linewidth}{!}{%
    \begin{tabular}{lllll}
        \toprule
        \multicolumn{1}{c}{\multirow{2}[3]{*}{\shortstack[c]{Model setup}}} & \multicolumn{4}{c}{Biodegradation} \\
        \cmidrule(lr){2-5}
        \multicolumn{1}{c}{}   & \multicolumn{1}{c}{ACC} & \multicolumn{1}{c}{SP} & \multicolumn{1}{c}{SN} & \multicolumn{1}{c}{AUROC} \\
        \midrule
        single GNN (training set) & 0.89 $\pm$ 0.03 & 0.94 $\pm$ 0.03 & 0.77 $\pm$ 0.10 & 0.96 $\pm$ 0.02  \\
        single GNN (validation set) & 0.86 $\pm$ 0.04 & 0.93 $\pm$ 0.03 & 0.72 $\pm$ 0.09 & 0.93 $\pm$ 0.03  \\
        single GNN (test set) & 0.85 $\pm$ 0.01 & 0.93 $\pm$ 0.03 & 0.66 $\pm$ 0.07 & 0.91 $\pm$ 0.01  \\
        ensemble of 40 GNNs (training/validation set) & 0.91 & 0.96 & 0.82 & 0.97  \\
        ensemble of 40 GNNs (test set) & 0.87 & 0.93 & 0.70 & 0.92  \\
        \bottomrule
    \end{tabular}
    }
    \caption{Model accuracies for ready biodegradation prediction: Accuracy is provided by accuracy score (ACC), specificity (SP), sensitivity (SN), all at a threshold value of 0.5, and area under the receiver operating characteristic curve (AUROC). For the single GNN model, the statistics are computed over 40 single models.}
    \label{tab:Prediction_BioDeg}
\end{table}

The averaged accuracy of the single GNN models amounts to 0.89 for the training set, 0.86 for the validation set, and 0.85 for the test set, indicating a high prediction quality, an observation that is further emphasized by specificity values exceeding 0.9 in case of all sets.
The sensitivity values are, however, visibly lower, e.g., a sensitivity of 0.66 is found for the test set.
The low sensitivity and high specificity point to the model's tendency of being cautious in classifying a molecule as readily biodegradable. 
Note that altering the threshold value on the predicted probability of a molecule being classified as readily biodegradable can be used to balance specificity and sensitivity.
We illustrate this trade-off with the receiver operating characteristic (ROC) curve below. 

We observe a slightly better prediction quality for the ensemble GNN.
For example, both the accuracy and sensitivity scores increase by 0.02 and 0.03, respectively.
The specificity with a value of 0.93, however, remains unchanged.
Overall, the ensemble thus only slightly enhances prediction performance in this example.

\begin{figure}[htbp]
    \centering
    \captionsetup[subfigure]{justification=centering}
	\begin{subfigure}[c]{0.5\textwidth}
		\centering
		\includegraphics[width=\textwidth]{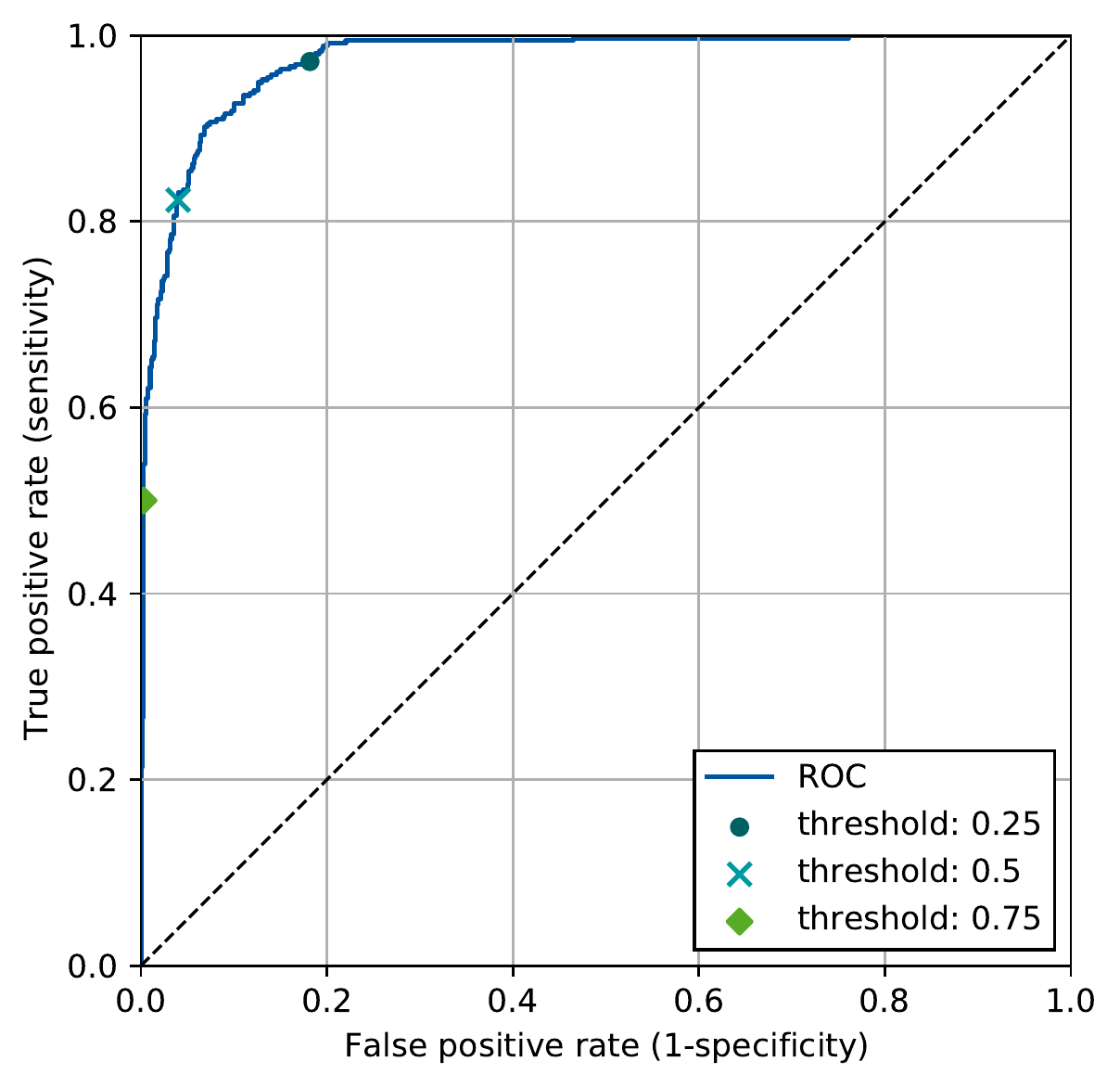}
		\subcaption{Training/validation set}
	\end{subfigure}%
	\begin{subfigure}[c]{0.5\textwidth}
		\centering
		\includegraphics[width=\textwidth]{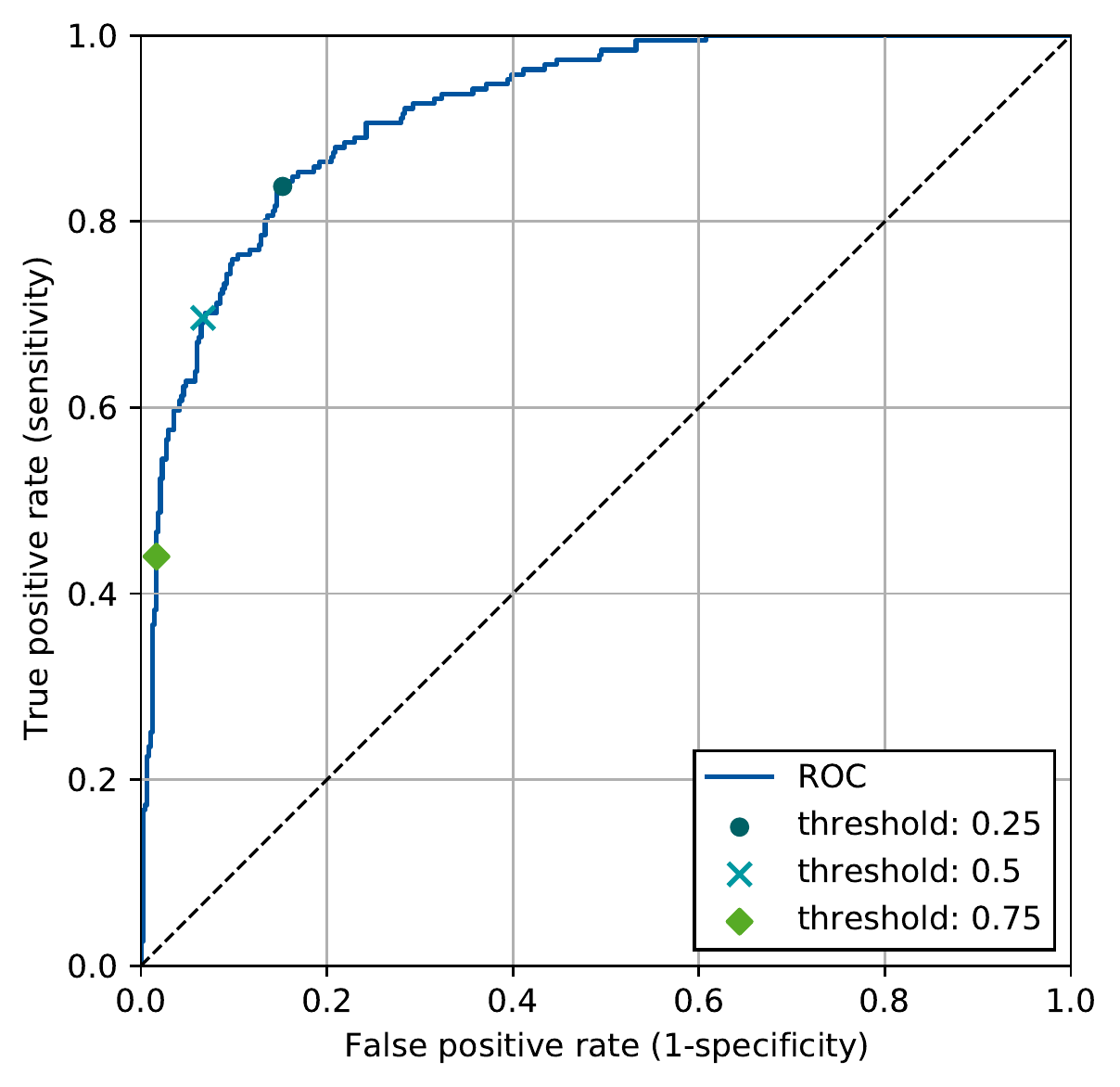}
		\subcaption{Test set}
	\end{subfigure}
	\caption{Receiver operating characteristic (ROC) curve in blue for GNN model ensembling.} 
	\label{fig:BioDeg_ROC}
\end{figure}

Figure~\ref{fig:BioDeg_ROC} shows the ROC curve for the training/validation set and the test set when using the GNN ensemble.
The ROC illustrates the true positive rate (sensitivity) versus the false positive rate ($1-$specificity) when varying the threshold probability value for deciding if a molecule is classified as readily biodegradable.
For example, if the threshold value is lowered, more molecules will be classified as biodegradable which increases the true positive rate but also the false positive rate.
A ROC curve with many high true positive and low false positive rate points, i.e., a curve located in the upper left corner of the graph, indicates a balanced classification model.
In Figure~\ref{fig:BioDeg_ROC}, we observe that the ROC curves for training/validation sets and test set are approaching that upper left region.
The good model performance is also reflected in the area under the receiver operating characteristic curve (AUROC) values (cf. Table~\ref{tab:Prediction_BioDeg}).
All values being larger than 0.9 is further evidence of an overall high prediction quality.
In fact, the GNN models for classifying molecules into readily biodegradable or not readily biodegradable yield similar prediction accuracy than state-of-the-art QSPR models~\cite{Mansouri2013, Goh.2018, Lee.2022}.

\section{Concluding remarks}\label{sec:ConclusionOutlook}
Graph neural networks (GNNs) have emerged as a powerful method for predicting physico-chemical properties of molecules allowing end-to-end learning from the molecular graph to the property of interest.
A principal advantage of GNNs is their straightforward applicability in situations where informative molecular descriptors relevant to the property of interest are not known and where the application of classical QSAR/QSPR modeling is therefore difficult.
We have illustrated the application of GNNs with standard architectures to property regression and classification tasks, i.e., the prediction of the normal boiling point and the prediction of biodegradability, respectively. 
Providing fast and high-quality molecular property predictions, GNNs provide a promising tool for in silico screening of molecules as well as computer-aided molecular design. 

GNNs constitute a fast moving research field, with novel GNN architectures being developed.
The novel architectures differ in many ways, for example, whether only representations of nodes~\cite{Hamilton.2017_Inductive, Gilmer.2017}, edges~\cite{Yang.2019}, or both~\cite{Kearnes.2016} are learned and what type of graph convolution and pooling function is applied. 
Recent GNNs for molecular property prediction also incorporate physical knowledge,
i.e., the message passing scheme is integrated with physical knowledge, e.g., SchNet~\cite{Schutt.2018}, DimeNet~\cite{Klicpera.2020}, MXMNet~\cite{Zhang2020_MXM}. 
This includes the incorporation of directional information, such as interatomic distances and angles between atom pairs, into the message function $M_l(\cdot)$.
Additional GNN extensions include applying a global message passing that enables direct interatomic information exchange between atoms that are not connected by a chemical bond, in contrast to local message passing as described above~\cite{Zhang2020_MXM}. 
Furthermore, GNNs have become more expressive from a graph theoretical perspective, i.e., more powerful in distinguishing graphs.
It was shown that GNNs are as powerful as the one-dimensional Weisfeiler-Lehmann algorithm in discriminating graphs~\cite{Xu.2018, Morris.2019}.
By including higher-order graph features~\cite{Morris.2019} or regular cell complexes~\cite{Bodnar.2021}, the expressiveness of GNNs can be increased.
For further insights on the theoretical foundations, we refer the interested reader to review articles~\cite{Xu.2018, Morris.2019, Morris.2021}.

Future GNN research will also need to focus on uncertainty quantification and interpretability of GNN predictions.
Next to the methodological advances, further application areas of GNNs in molecular property prediction will be explored, as well as beyond. 
In chemical engineering, for instance, also process flow sheets or reaction networks can be represented as graphs, suggesting another huge potential for GNN applications~\cite{schweidtmann2021machine}. 

\section*{Acknowledgements}
This work was funded by the TU Delft AI Labs Programme.
This work was funded by the Deutsche Forschungsgemeinschaft (DFG, German Research Foundation) – 466417970 – within the Priority Programme ``SPP 2331: Machine Learning in Chemical Engineering''.
MD received funding from the Helmholtz Association of German Research Centres.

\bibliographystyle{unsrt}
\bibliography{bibliography}
\end{document}